# More on alleged extraordinariness of graphene as a nonlinear optical material

Jacob B Khurgin,

*Johns Hopkins University, 3400 N. Charles Street, Baltimore MD 21218  USA*

*jakek@jhu.edu*

Graphene may be a fascinating material but at this time there exists no experimental or theoretical evidence that it can improve performance of practical nonlinear optical devices.

In 2014 a paper had been published in APL [1] in which I had the audacity to show that graphene does not possess nonlinear properties that are superior to other nonlinear material in any meaningful way. In his most recent arXiv contribution [2] S. A. Mikhailov challenges this opinion claiming that graphene does possess nonlinear properties orders of magnitude superior to those of other material although no experimental evidence in support of this opinion is given.

The centerpiece of [2] is Equation (5) which in context of the issue discussed here is (a) not relevant (b) not correct and (c) evaluated using erroneous data. This equation is not relevant because the subject of Ref [1] is the nonlinear index of refraction and ensuing phase change. Phase change requires real nonlinear susceptibility $\chi^{(3)}$ (and therefore imaginary nonlinear conductivity $\sigma^{(3)} = i\omega\varepsilon_0\chi^{(3)}$ . When one operates above the transparency region ($\hbar\omega > 2E_F$) as supposed in [2] the nonlinear conductivity $\sigma^{(3)}(\omega,\omega,-\omega)$ as evidenced from Eq(7) in [2] is real, hence the process describes change (saturation) of the absorption and not the phase change (retardation) considered in [1]

Now, if we are to deviate from the main topic of [1], and to deal with absorption saturation, expression (5) in [2] happens to be incorrect, because when absorption occurs the real carriers get excited, hence their scattering lifetimes must be involved and one should not use perturbative expression (3). Instead one can easily derive the well-known expression describing the saturation of absorption as

$$\frac{\sigma^{(3)}}{\sigma^{(1)}} = \frac{e^2 P_{cv}^2}{\hbar^2 m_0^2 \omega^2} T_{coh} T_{scat},$$

where $P_{cv}$ is the matrix element of momentum between conduction and valence bands and $m_0$ is free electron mass. For graphene $P_{cv}/m_0 = v_F = 10^8 cm/s$. For GaAs and wide range of other III-V or II-VI semiconductors $E_P = 2P_{cv}^2/m_0 \sim 20-30 eV$ [3] and therefore $P_{cv}/m_0 \sim .95-1.1\times 10^8 cm/s$ i.e. *comparable to that of graphene*. This should come as no surprise given that two dipole allowed transitions in the same spectral range are expected to have the same oscillator strength.

Furthermore, $T_{coh}$ in the above expression is the *coherence* time, and *not the interband r*elaxation time as claimed in the comment [2] . hence $T_{coh}^{-1}$ defines the width of spatial hole burned in the absorption spectrum and $T_{scat}^{-1}$ is the *intraband* scattering rate of carriers which determines the temporal response of the nonlinearity. Both scattering and coherence times are largely defined by the *intraband* electron-electron scattering as well as phonons scattering. This processes occur on a scale of *20-30fs* [4-7] which is about 12-15 times less than $330 fs$ *interband* time used in [2]. Therefore the values of $\sigma^{(3)}$ reported in [2] using $\gamma = 2meV$ should be estimated using $\gamma = 20-30 meV$ or less and thus be reduced by at the very least two orders of magnitude, much closer to the estimate given in [1].

Now, when $\sigma^{(3)}E^2$ approaches $\sigma^{(1)}$ the absorption saturation takes place and, and one can estimate the saturation power density as $I_{sat} = E^2/2\eta_0$ where $\eta_0 = 377\Omega$. If one is to believe [2] the saturation occurs when the optical field is 10KV/cm which means that $I_{sat} \sim 100 kW/cm^2$. Experimental data [8,9] supported by theory [8] clearly shows that saturation takes place at $I_{sat} \sim 1-10G W/cm^2$, i.e *4-5 orders of magnitude higher than predicted in [2],* roughly in line with any other saturable absorber, such as SESAM (which has been and remains the device of choice in mode locked lasers). Of course, *in graphene as in any other* *material* one can reduce the saturation intensity [10] by increasing the scattering time $T_{scat}$ but this will also slow down the effect, and here as in [1] *we are only interested in ultrafast nonlinearity*

After discussing interband nonlinearities the author of comment [2] shifts to intraband nonlinearities and claims that nonlinearity is very high at f=1THz which is true, but one cannot compare apples with oranges –nonlinearity increases super linearly with the decrease in frequency. At 1THz and below the competition to graphene is not conventional nonlinear materials but rather electronic devices, such as Schottky diodes and HEMT transistors where nonlinearity is immeasurably higher than in graphene. At higher frequencies such as 30THz nonlinear conductivity may be still be high, *but this nonlinear conductivity quickly saturates and one cannot achieve $2\pi$ phase shift* as discussed in [1] at length. At any rate, any narrow band semiconductor with strong conduction band non-parabolicity will show the effects comparable in strength to graphene. Intersubband transitions in semiconductor quantum wells show higher nonlinearity than graphene and they (unlike graphene) in fact had been successfully used for THz generation with good efficiency [11]

To draw this reply to a close, I express no intention to contest the concluding statement made in [2] that "the nonlinear graphene optics and electrodynamics is a promising and encouraging field of research" as I uphold the prevalent opinion that any scientist should feel no constraints to explore just about anything that his or her program manager deems reasonable. My point is only to reiterate that as any practitioner of nonlinear optics knows, the strength of nonlinear effects depends on the very few material parameters, namely the oscillator strength (optical transition dipole), detuning from resonance, broadening, and in some (above the bandgap) cases also on densities of states and relaxation times. If one compares graphene with a semiconductor QW made from any direct bandgap semiconductor including organic ones (perovskites) or with other 2D materials ($WSe_2$) these parameters are all in the same range, hence the nonlinear properties of graphene are also comparable to most other decent nonlinear materials as numerous experiments have shown. As I have made clear in the original article [1], graphene may indeed find its proper place in nonlinear optical applications, though not because of the intrinsically far superior performance, but due to more practical reasons such as cost, ease of fabrication, or tunability over wide spectral region.